\documentclass[twocolumn,printnumbers,amsmath,amssymb]{revtex4}

\usepackage{graphicx}
\usepackage{dcolumn}
\usepackage{bm}
\usepackage{supertabular}

\begin{document}


\title{
Dynamical dimer method for the determination of transition states with
ab initio molecular dynamics
}

\author{Alexander Poddey}%
 \email{alexander.poddey@tu-clausthal.de}
\author{Peter E. Bl\"ochl}%
 \email{peter.bloechl@tu-clausthal.de}
\affiliation{%
Clausthal University of Technology, Institute of Theoretical Physics\\
Leibnizstrasse 10, D-38678 Clausthal-Zellerfeld, Germany
}%

\date{\today}

\begin{abstract}
  A dynamical formulation of the dimer method for the determination of
  transition states is presented. The method is suited for ab-initio
  molecular dynamics using the fictitious Lagrangian formulation. The
  method has been applied to the con-rotatory ring opening of
  chloro-cyclo-butadiene, an example, where the application of the drag
  method is problematic.
\end{abstract}

\maketitle

\section{\label{sec:intro}Introduction}
The concept of the transition state has a fundamental role for the
prediction of rate constants of materials processes.  The transition
state is the lowest point on the energy barrier separating two metastable
states representing the initial and final state of a process.  The
energy difference between the initial state and the transition state
is the activation energy. The curvatures of the potential energy
surface at the initial state and the transition state provides an
estimate of entropic effects\cite{ts1,ts2}. A molecular dynamics
simulation starting from the transition state allows one to construct
the dynamical trajectory of a reaction in the low-temperature limit.
The extension of this latter approach to finite temperatures is
straightforward \cite{keck62}.

The determination of transition states is however substantially more
difficult than the determination of minima of the total energy
surface.  A large number of methods have been invented and refined
over the last decades (see e.g. the Refs.
\onlinecite{fts1,fts2,fts3,fts4,fts5,fts6,fts7,fts8,fts9,fts10,fts11,fts12}
and the review articles \cite{ftsover,ftsover2}). 

Many of these
methods require the determination of the second derivatives of the
total energy surface. While forces, that is first derivatives of the potential energy
surface, are readily accessible from density functional calculations,
second derivatives are not, at least not without substantial effort.

An overview of methods which allow to determine transition states
solely from force information can be found in \cite{ftsover}.
The most simple approach is the drag method, where a one-dimensional
constraint forces the system across a reaction barrier. Widely used
are also ``chain of states'' algorithms such as the nudged elastic
band (NEB) \cite{fts7,fts8,fts9} or the string method
\cite{fts11,fts12}. 

Intermediate between these extremes is the dimer method.
\cite{dimer1}. It couples two instances of the system, which we will
call monomers. The monomers have a specified distance in configuration
space. If the system follows the forces, after inverting of the force parallel
to the dimer, it evolves to a saddle point of first order, that
is the transition state.

The original dimer method \cite{dimer1}
applied a two step algorithm. In a first step, the potential energy of
the dimer-construct is minimized, excluding parallel motion, following a conjugate gradient
approach. Then a large step is taken in the direction parallel to the
dimer axis. Recently, an improvement has been suggested which
minimizes the number of gradient evaluations \cite{dimer2}.

In the context of fictitious-Lagrangian approach to first-principles
molecular dynamics (FPMD) \cite{cp}, such a two-step procedure is not
feasible, because the electronic wave functions must be able to follow
the atomic motion adiabatically. The fictitious Lagrangian approach requires a larger
number of time steps. However the computational cost per time step are minimal.
While a first-principles string approach has been published recently
\cite{stringcar}, the dimer method has not been examined in the
framework of ab initio molecular dynamics.

In the present study, we introduce a dynamical formulation of the
dimer method. Starting from an extended Lagrangian for the dimer in
the configuration space with doubled dimensionality, we derive
equations of motion, that, in the presence of dissipation, converge at
transition states of first order. An analysis of the trajectories of
the dimer close to the stable points provides information on the
stability and gives guidance for improving the performance of the
method.  The method has been implemented in the CP-PAW code and has
been applied to the con-rotatory ring-opening of chloro-cyclo-butene, an
example where the conventional drag method fails unless special
precautions are taken.

The paper is organized as follows: In section \ref{sec:flf} we define the extended Lagrangian for the
dimer motion and derive the equations of motion. In
section~\ref{sec:lsa} we analyze stability and the dynamics in the
proximity of the stable points. Section \ref{sec:numint} is devoted to
the discretization the equations of motion and the aspects of the
actual implementation. In the final section \ref{sec:app}, the method
is applied to the con-rotatory ring opening of cyclo-butene as
practical example and compared to the drag method.


\section{\label{sec:flf}Lagrangian and equations of motion}
In this section we define the Lagrangian underlying the present work
and derive the equations of motion. We begin by defining our notation and we introduce a convenient set of
coordinates:

The dimer consists of two points in configurational space separated by
a fixed distance.  We use mass-weighted coordinates $\mathbf{x}_i$
defined as
\begin{eqnarray}
\mathbf{x}_i := \mathbf{\underline{m}}^{\frac{1}{2}}\mathbf{R}_i ,
\label{eq:massweightedcoordinates}
\end{eqnarray} 
where the vectors $\mathbf{R}_i$ with $i=1,2$ are the positions of the
two monomers forming the dimer in configurational space. For
a system with $N$ atoms each vector has $3N$ dimensions.
In addition, each monomer has its own set of electronic wave
functions. The mass
matrix $\mathbf{\underline{m}}$ is diagonal and the masses of the respective
atoms are located on the main diagonal.

It is convenient to introduce a variable transformation into
center-of-gravity coordinates $\mathbf{Y}_1$ and relative coordinates
$\mathbf{Y}_2$.
\begin{eqnarray}
\mathbf{Y}_{1} & = & \frac{\mathbf{x}_{1}+\mathbf{x}_{2}}{2}\\
\mathbf{Y}_{2} & = & \mathbf{x}_{1}-\mathbf{x}_{2}
\end{eqnarray}
Loosely speaking $\mathbf{Y}_1$ describes the mean structure of the
molecule, while $\mathbf{Y}_2$ describes the orientation of the dimer.

A dimer has three basic types of motion: motion parallel to the dimer
axis ($\mathbf{v}_\parallel$), motion perpendicular to the axis
($\mathbf{v}_\perp$) and rotation about the center of gravity
($\mathbf{v}_\circ$).  In the following, we divide the velocities into
the contributions from these basic types of motion. The corresponding
velocities are
\begin{eqnarray}
\mathbf{v}_{\circ}&=& \mathbf{\dot{Y}}_2
=\mathbf{\dot{x}}_{1}-\mathbf{\dot{x}}_{2}
\label{eq:vcirc}
\\
\mathbf{v}_{\parallel} & = &\frac{\mathbf{Y}_2\otimes\mathbf{Y}_2}{d^2}
\mathbf{\dot{Y}}_1 \nonumber \\ 
&=& \left(\mathbf{x}_{1}-\mathbf{x}_{2}\right)
\frac{\left(\mathbf{x}_{1}-\mathbf{x}_{2}\right)
\left(\mathbf{\dot{x}}_{1}+\mathbf{\dot{x}}_{2}\right)}{2d^{2}}
\label{eq:vparallel}
\\
\mathbf{v}_{\perp} & = &
\left(1-\frac{\mathbf{Y}_2\otimes\mathbf{Y}_2}{d^2}\right)
\mathbf{\dot{Y}}_1
=
 \frac{\mathbf{\dot{x}}_{1}+\mathbf{\dot{x}}_{2}}{2}-\mathbf{v}_{\parallel}
\label{eq:vperp}
\end{eqnarray}
The dot denotes the time derivative, and $d$ is the so called 'dimer
distance'
$d=\sqrt{\mathbf{Y}_2^2}=\sqrt{\left(\mathbf{x}_{1}-\mathbf{x}_{2}\right)^2}$.
The symbol $\otimes$ refers to the outer product defined by
$\left(\mathbf{a}\otimes\mathbf{b}\right)\mathbf{c}
=\mathbf{a}\left(\mathbf{b}\cdot\mathbf{c}\right)$.  Note, that
$\frac{\mathbf{Y}_2\otimes\mathbf{Y}_2}{d^2}$ is an operator that
projects a vector onto the dimer axis.

Now we define our theory by setting up a Lagrangian function in the
relative and center-of-gravity coordinates
\begin{eqnarray}
\mathcal{L}(\mathbf{Y}_1,\mathbf{Y}_2,\mathbf{\dot{Y}}_1,\mathbf{\dot{Y}}_2) & = &M_{\circ}\frac{1}{4}\mathbf{v}_{\circ}^{2}
+M_{\perp}\mathbf{v}_{\perp}^{2}
+M_{\parallel}\mathbf{v}_{\parallel}^{2}\nonumber\\ 
&&-V\left(\mathbf{\underline{m}}^{-\frac{1}{2}}\left[\mathbf{Y}_1+\frac{1}{2}\mathbf{Y}_2\right]\right)\nonumber\\
&&-V\left(\mathbf{\underline{m}}^{-\frac{1}{2}}\left[\mathbf{Y}_1-\frac{1}{2}\mathbf{Y}_2\right]\right)\nonumber\\ 
&&-\bar{\lambda}\left[\mathbf{Y}_{2}^2-d^{2}\right]
\label{eq:lagrangefunction}
\end{eqnarray}
The potential energy is described by the potential energy surface
$V(\mathbf{R})$.  The velocities $\mathbf{v}_\circ$,
$\mathbf{v}_\perp$ and $\mathbf{v}_\parallel$ are to be treated as
functions of $\mathbf{Y}_1$ and $\mathbf{Y}_2$ and their time
derivatives $\mathbf{\dot{Y}}_1$ and $\mathbf{\dot{Y}}_2$ as defined
in Eqs.~\ref{eq:vcirc},\ref{eq:vparallel},\ref{eq:vperp}.

We introduced three scale factors $M_\circ$, $M_\parallel$ and
$M_{\perp}$, that allow us to scale the masses for the three types of
motions independently, that will later be used to accelerate
convergence. More importantly, by choosing a negative value of
$M_\parallel$, the motion along the dimer direction is inverted, so
that the dimer climbs up the barrier.

If the scale factors are equal to one, the conventional kinetic
energy is recovered, that is
\begin{eqnarray}
\frac{1}{4}\mathbf{v}_\circ^2+\mathbf{v}_\parallel^2+\mathbf{v}_\perp^2
=\frac{1}{2}\mathbf{\dot{R}}_1\mathbf{m}\mathbf{\dot{R}}_1
+\frac{1}{2}\mathbf{\dot{R}}_2\mathbf{m}\mathbf{\dot{R}}_2
\end{eqnarray}

Using the method of Lagrange multipliers, a term has been introduced
to describe the dimer-distance constraint
\begin{eqnarray}
\mathbf{Y}_{2}^2=d^{2}
\end{eqnarray}
which ensures that the dimer distance in mass-weighted coordinates is
equal to $d$. The corresponding Lagrange multiplier is denoted by $\bar{\lambda}$.

In addition to the Lagrangian we also define a Rayleigh's dissipation
function $\mathcal{D}$
 \begin{eqnarray}  
\mathcal{D}&=&\gamma_{\circ}M_{\circ}\frac{1}{4}\mathbf{v}_{\circ}^2+
\gamma_{\parallel}M_{\parallel}\mathbf{v}_{\parallel}^2
+\gamma_{\perp}M_{\perp}\mathbf{v}_{\perp}^2\,.
\label{eq:rayleigh}
\end{eqnarray}  
which allows us to introduce dissipation in a consistent manner.

The Euler-Lagrange equation for the Lagrangian of
Eq.~\ref{eq:lagrangefunction} and the Rayleigh's Dissipation function
from Eq.~\ref{eq:rayleigh} are obtained from
\begin{eqnarray}
 \frac{d}{dt}\frac{\partial\mathcal{L}}{\partial\dot{Y}_{i,n}}
-\frac{\partial\mathcal{L}}{\partial Y_{i,n}}
+\frac{\partial\mathcal{D}}{\partial\dot{Y}_{i,n}} 
= 0\,,
\label{eq:elwq} 
\end{eqnarray}
where the index $i\in\{1,2\}$ labels the two monomers, and
$n\in\{1,3N\}$ labels the coordinate in configuration space.  

The resulting equations of motion have the form
\begin{eqnarray}
\left(1-\frac{\mathbf{Y}_2\otimes\mathbf{Y}_2}{d^2}\right)
M_\perp\left(\ddot{\mathbf{Y}}_1+\gamma_\perp\mathbf{\dot{Y}}_1\right)
\nonumber\\
+\left(\frac{\mathbf{Y}_2\otimes\mathbf{Y}_2}{d^2}\right)
M_\parallel\left(\ddot{\mathbf{Y}}_1+\gamma_\parallel\mathbf{\dot{Y}}_1\right)\nonumber \\
-\frac{1}{2}\mathbf{m}^{-\frac{1}{2}}\left(\mathbf{F}_1+\mathbf{F}_2\right)
\nonumber\\
+\left(M_\parallel-M_\perp\right)\left[
\frac{d}{dt}\left(\frac{\mathbf{Y}_2\otimes\mathbf{Y}_2}{d^2}\right)
\right]\mathbf{\dot{Y}}_1&=&0
\label{eq:eqm1}
\end{eqnarray}

and
\begin{eqnarray}
 &&M_{\circ}\ddot{\mathbf{Y}}_{2}-\underline{\mathbf{m}}^{-\frac{1}{2}}\left(\mathbf{F}_{1}-\mathbf{F}_{2}\right)+4\lambda\mathbf{Y}_{2}
\nonumber\\
&&
+4\left(M_{\perp}-M_{\parallel}\right)\frac{\mathbf{Y}_{2}\mathbf{\dot{Y}}_{1}}{d^{2}}
\mathbf{\dot{Y}}_{1}
+\gamma_{\circ}M_{\circ}\mathbf{\dot{Y}}_{2}=0
\label{eq:eqm2}
\end{eqnarray}
Here we have used the forces acting on the monomers defined as
\begin{eqnarray}
\mathbf{F}_i:=-\left.\mathbf{\nabla}\right|_{\mathbf{R}_i} V
\end{eqnarray}

In equation (\ref{eq:eqm2}) we absorbed all terms which lead to a
force parallel to the dimer axis into the constraint force by
redefining the Lagrange multiplier.  Thus the variable $\lambda$ used
in Eq.~\ref{eq:eqm2} differs from the Lagrange multiplier
$\bar{\lambda}$ in Eq.~\ref{eq:lagrangefunction}.

Furthermore, because the dimer system moves on the hyperplane with
constant dimer distance, we simplified the final equations by using
\begin{eqnarray}
\mathbf{Y}_2^2=d^2
\end{eqnarray}
and
\begin{eqnarray}
\frac{\partial}{\partial t}\mathbf{Y}_2^2=0\,.
\end{eqnarray}
Equation (\ref{eq:eqm1}) describes the motion of the center of gravity of
the dimer, which is related to the structure of the molecule. Equation 
(\ref{eq:eqm2}) describes the orientational motion of the dimer.


\section{\label{sec:lsa}Local stability analysis}
For $M_{\circ}=M_{\parallel}=M_{\perp}=1$, the Lagrange function
(\ref{eq:lagrangefunction}) describes a physical system of two
masspoints 
moving in a potential $V$ under the influence of the constraint force
that keeps the dimer distance invariant. With positive friction
factors $\gamma_\circ$, $\gamma_\parallel$ and $\gamma_\perp$ the
dimer will come to rest with the center-of-gravity coordinate next to
a local minimum. The dimer axis will be aligned nearly parallel to the
lowest vibrational eigenmode. The above statements are exactly fulfilled
in the limit of vanishing dimer distance.

If we choose a negative value of $M_{\parallel}$, the motion will
become unstable near local minima. Instead, the dimer will be
attracted by transition states of first order.  A transition state of
first order is characterized by the presence of exactly one eigenmode
with an imaginary frequency.

These properties of the dynamics have been derived from the following
local stability analysis. First we determine the stationary points for
the dimer dynamics. Then we investigate the dynamics in the
neighborhood of those stationary points.

For this analysis we can replace the potential energy surface by its
truncated Taylor expansion at a given point
$\mathbf{R}^{(0)}=\mathbf{m}^{-\frac{1}{2}}\mathbf{x}^{(0)}$, namely
\begin{eqnarray}
V(\mathbf{R})
&=&V^{(0)}-\mathbf{F}^{(0)}\mathbf{R}\nonumber\\
&&+\frac{1}{2}(\mathbf{R}-\mathbf{R}^{(0)})\mathbf{m}^\frac{1}{2}
\mathbf{D}\mathbf{m}^\frac{1}{2}(\mathbf{R}-\mathbf{R}^{(0)})
\label{eq:taylorpotential}
\end{eqnarray}
with the dynamical matrix $\mathbf{D}$ defined as
\begin{eqnarray}
D_{i,j}=\frac{1}{\sqrt{m_{i,i}}}
\left.\frac{\partial^2V}{\partial R_i\partial R_j}\right|_{\mathbf{R}^{(0)}}
\frac{1}{\sqrt{m_{j,j}}}
\end{eqnarray}
From Eq.~\ref{eq:taylorpotential} and
Eq.~\ref{eq:massweightedcoordinates}, we obtain the forces
\begin{eqnarray}
\mathbf{F}(\mathbf{x})=\mathbf{F}^{(0)}-\mathbf{m}^{\frac{1}{2}}\mathbf{D}(\mathbf{x}-\mathbf{x}^{(0)})
\label{eq:linforce}
\end{eqnarray}

\subsection{Stationary points}
\label{sec:stationarypoints}
If we insert the condition for a stationary point
$\mathbf{\dot{Y}}_{i}=\mathbf{\ddot{Y}}_{i}=0$ in the equation of
motion Eq.~\ref{eq:eqm1}, we find that the forces for the two monomers
at the stationary point are antiparallel and of the same magnitude.
Using the Taylor expansion Eq.~\ref{eq:linforce} in the equation of
motion Eq.~\ref{eq:eqm1}, we find that the center-of-gravity of the
dimer $\mathbf{Y}_1$ lies at a stationary point of the potential, if
the dimer is stationary, that is
\begin{eqnarray}
\mathbf{F}(\mathbf{Y}_1)=0
\end{eqnarray}

Similarly we obtain from Eq.~\ref{eq:eqm2} and Eq.~\ref{eq:linforce}
\begin{eqnarray}
\mathbf{D}\mathbf{Y}_2=-4\lambda\mathbf{Y}_2
\label{eq:eigenvalue}
\end{eqnarray}
Thus the dimer axis points along an eigenvector of the dynamical
matrix.

In conclusion we find that the dimer dynamics is stationary, when its
center of gravity lies at an extremum or a saddle point of the
potential energy surface and when, in addition, the dimer axis points
along one of the vibrational eigenmodes.

\subsection{\label{sec:}Linearized equation of motion}
Without loss of generality, we choose in the following a coordinate system for which the
stationary point of the potential energy surface lies at the
origin. Secondly, we denote the eigenvectors of the dynamical matrix as
$\mathbf{e}_i$, with $\mathbf{e}_i\mathbf{e}_j=\delta_{i,j}$, and the
corresponding eigenvalues with $\omega_i^2$. The eigenvector, which is
parallel to the dimer axis is denoted by $\mathbf{\bar{e}}$ and the
corresponding eigenvalue is denoted by $\bar{\omega}^2$. Thus, with
Eq.~\ref{eq:eigenvalue}, we can identify the Lagrange multiplier as
$\lambda=-\frac{1}{4}\bar{\omega}^2$.

Linearization of the equations of motion Eqs.~\ref{eq:eqm1} and
\ref{eq:eqm2} about $\mathbf{Y}_1=\mathbf{0}$ and
$\mathbf{Y}_2=\mathbf{\bar{e}}d$ with $\mathbf{F}^{(0)}=0$ yields the
following equations for the deviation $\delta\mathbf{Y}_1(t)$ and
$\delta\mathbf{Y}_2(t)$ from the stationary point
\begin{eqnarray}
\left(1-\mathbf{\bar{e}}\otimes\mathbf{\bar{e}}\right)
M_\perp\left(\delta\ddot{\mathbf{Y}}_1
+\gamma_\perp\delta\dot{\mathbf{Y}}_1\right)&&
\nonumber\\
+\left(\mathbf{\bar{e}}\otimes\mathbf{\bar{e}}\right)
M_\parallel\left(\delta\ddot{\mathbf{Y}}_1
+\gamma_\parallel\delta\dot{\mathbf{Y}}_1\right)
+\mathbf{D}\delta\mathbf{Y}_1
&=&0
\label{eq:linear1}
\\
M_\circ\delta\ddot{\mathbf{Y}}_2+\mathbf{D}\delta\mathbf{Y}_2
+\bar{\omega}^2\delta\mathbf{Y}_2
+M_\circ\gamma_\circ\delta\dot{Y}_2&=&0
\label{eq:linear2}
\end{eqnarray}

We project the second equation ,Eq.~\ref{eq:linear2}, onto the
eigenvectors $\mathbf{e}_i$ and obtain: 
\begin{eqnarray}
M_\circ(\mathbf{e}_i\delta\ddot{\mathbf{Y}}_2)
+(\omega_i^2-\bar{\omega}^2)(\mathbf{e}_i\delta\mathbf{Y}_2)
+M_\circ\gamma_\circ(\mathbf{e}_i\delta\dot{Y}_2)=0
\end{eqnarray}
Note, that the motion along
$\mathbf{\bar{e}}$ is simple due to the distance constraint.
The frequency of the variable
$\mathbf{e}_i\mathbf{Y}_2$ is therefore
\begin{eqnarray}
\omega_\circ=\pm\sqrt{\frac{\omega_i^2-\bar{\omega}^2}{M_\circ}}
\label{eq:freqcirc}
\end{eqnarray}
We see that, for a positive mass $M_\circ$, the dynamics is stable
only, if $\bar{\omega}^2$ is the lowest eigenvalue of the dynamical
matrix.  Thus the dimer axis will always orient along the eigenvector
with the lowest eigenvalue.  

Now, we project Eq.~\ref{eq:linear1} onto the eigenvector
$\mathbf{\bar{e}}$ of the dynamical matrix
\begin{eqnarray}
M_\parallel(\mathbf{\bar{e}}\delta\ddot{\mathbf{Y}}_1)
+\bar{\omega}^2(\mathbf{\bar{e}}\delta\mathbf{Y}_1)
+M_\parallel\gamma_\parallel(\mathbf{\bar{e}}\delta\dot{\mathbf{Y}}_1)
=0
\label{eq:linear1parallel}
\end{eqnarray}
which yields the translational motion of the dimer along the dimer axis.
This equation will be responsible for the ascent to a saddle
point. The eigenfrequency of the variable $\mathbf{\bar{e}}\mathbf{Y}_1$ is
\begin{eqnarray}
\omega_\parallel=\pm\frac{\bar{\omega}}{\sqrt{M_\parallel}}
\label{eq:freqpara}
\end{eqnarray}
For a negative mass $M_\parallel$, which is the choice for a
transition state search, Eq.~\ref{eq:linear1parallel} is stable only
if the dimer is oriented along an unstable vibrational mode of the
potential energy surface, that is with the dimer oriented along an
eigenvector $\mathbf{\bar{e}}$ with an imaginary frequency.

Now, we project Eq.~\ref{eq:linear1} onto the eigenvectors
$\mathbf{e}_i$ with $\mathbf{e}_i\perp\mathbf{\bar{e}}$ and we obtain
\begin{eqnarray}
M_\perp(\mathbf{e}_i\delta\ddot{\mathbf{Y}}_1)
+\omega_i^2(\mathbf{e}_i\delta\mathbf{Y}_1)
+M_\perp\gamma_\perp(\mathbf{e}_i\delta\dot{\mathbf{Y}}_1)
=0
\label{eq:linear1perp}
\end{eqnarray}
which represents the translational motion perpendicular to the dimer
axis. The translational motion is related to an optimization of the
atomic structure.  The frequency of the variable
$\mathbf{e}_i\mathbf{Y}_1$ with $\mathbf{e}_i\perp\mathbf{\bar{e}}$ is
\begin{eqnarray}
\omega_\perp=\frac{\omega_i}{\sqrt{M_\perp}}
\label{eq:freqperp}
\end{eqnarray}

From Eqs.~\ref{eq:freqcirc}, \ref{eq:freqpara} and \ref{eq:freqperp} we
find that the dimer with positive $M_\circ$ and $M_\perp$ and negative
$M_\parallel$ will only come to rest at transition states of first
order, that is, if $\bar{\omega}^2<0$ and if all other eigenvalues are
positive. In that case the dimer axis points along the eigenvector
corresponding to the imaginary frequency, that is, across the saddle
point.

\subsection{\label{sec:convergence}Masses and Frictions}
\begin{figure}[h!]
\begin{center}
\includegraphics[width=8cm]{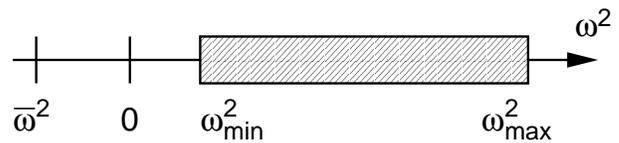}
\end{center}
\caption{Sketch of the eigenvalue spectrum of the dynamical matrix at
the saddle point of first order with the definition of $\omega_{min}$
and $\omega_{max}$.}
\label{fig:spectrum}
\end{figure}

We will later solve the equations of motion \ref{eq:eqm1}, \ref{eq:eqm2} 
with the Verlet algorithm. The
Verlet algorithm for a harmonic oscillator with a frequency $\omega$
becomes unstable, if the time step $\Delta$ used for the discretization of the
equation of motion is smaller than $2/\omega$. The frequency of
the discretized motion is accurate to within 1~\% if
$\omega\Delta<\frac{\pi}{5}$. Thus, it is important to understand the
vibrational spectrum, shown schematically in Fig.~\ref{fig:spectrum},
of the motion in the potential.

Thus, the stability limit requires us to choose the masses as
\begin{eqnarray}
M_\parallel&<&\frac{\bar{\omega}^2\Delta^2}{8}
\\
M_\perp&>&\frac{\omega_{max}^2\Delta^2}{8}
\\
M_\circ&>&\frac{(\omega_{max}^2-\omega_{min}^2)\Delta^2}{4}
\end{eqnarray}
In order to minimize the number of time steps, it is desirable to
choose the masses close to these limits. 

Let us now obtain an \textit{a priori} estimate of these limits: (1)
For practical purposes, we can make the assumption that
$\omega_{min}<<\omega_{max}$ and thus we set $\omega_{min}$ to zero.
(2) The highest vibrational mode is that of H$_2$, with a frequency of
$4400$ cm$^{-1}$. Frequencies for bond-stretch vibrations are of the
order $1000$ cm$^{-1}$. These
numbers are reasonable upper estimates for $\omega_{max}$.  (3) The
absolute value of the imaginary frequency at the saddle point is
typically in the range of the high-lying real-frequency modes, that is
comparable to $\omega_{max}$. Hence $\bar{\omega}^2\approx
-\omega_{max}^2$. With these assumptions we obtain the following,
recommended values for the masses 
\begin{eqnarray}
M_\parallel&=&-\kappa\Delta^2
\\
M_\perp&=&\kappa\Delta^2
\\
M_\circ&=&2\kappa\Delta^2
\end{eqnarray}
where $\kappa=\left(6\cdot10^{-3} ~\mathrm{a.u.}\right)^2$ is a recommended constant based on
the H$_2$ vibration.

Let us now turn our attention to the friction values. The goal is to
reach an optimum convergence at the stationary point of the dimer.
For a damped harmonic oscillator 
\begin{eqnarray}
m\ddot{x}=-m\omega_0^2x-m\gamma x
\end{eqnarray}
the fastest decay rate as function of the friction is obtained for
critical damping, that is in between the oscillatory and the
over-damped regime. Critical damping is obtained for a friction of
$\gamma=2\omega_0$. Thus we choose
\begin{eqnarray}
&\gamma_\parallel&=2\sqrt{\frac{\bar{\omega}^2}{M_\parallel}}
\label{eq:optfricpara}
\\
2\sqrt{\frac{\omega_{min}^2}{M_\perp}}
<&\gamma_\perp&<
2\sqrt{\frac{\omega_{max}^2}{M_\perp}}
\label{eq:optfricperp}
\\
2\sqrt{\frac{\omega_{min}^2-\bar{\omega}^2}{M_\circ}}
<&\gamma_\circ&<
2\sqrt{\frac{\omega_{max}^2-\bar{\omega}^2}{M_\circ}}
\label{eq:optfricrot}
\end{eqnarray}
Note, that a too high friction freezes out those modes in the
over-damped regime. Thus it is usually better to use a friction near the
lower bound of the sensible regime.

\section{\label{sec:numint}Numerical integration of the equations of motion}

\subsection{Discretization of the equations of motion}
The equations of motion Eqs.~\ref{eq:eqm1},\ref{eq:eqm2} are nonlinear
in the velocities. This leads to a nonlinear equation for the
discretized equation of motion. We tackle the problem by iterating on
the nonlinear terms in the velocities.

We set up the discretized equations of motion, while treating the
terms nonlinear in the velocities as an abstract force. 
\begin{eqnarray}
\mathbf{G}_1&:=&-\left(M_\parallel-M_\perp\right)\left[
\frac{d}{dt}\left(\frac{\mathbf{Y}_2\otimes\mathbf{Y}_2}{d^2}\right)
\right]\mathbf{\dot{Y}}_1
\\
\mathbf{G}_2&:=&4\left(M_\parallel-M_\perp\right)
\frac{\mathbf{Y}_2\dot{\mathbf{Y}}_1}{d^2}\mathbf{\dot{Y}}_1
\end{eqnarray}

The equations of motion are discretized according to the Verlet
algorithm:
\begin{eqnarray}
\dot{y}&\rightarrow& \frac{y(+)-y(-)}{2\Delta}
\\
\ddot{y}&\rightarrow& \frac{y(+)-2y(0)+y(-)}{\Delta^2}
\end{eqnarray}
where $\Delta$ is the discretization time step, and where we used the short-hand notation
\begin{eqnarray}
y(+)&=&y(t+\Delta) \\
dy(0)&=&y(t) \\
y(-)&=&y(t-\Delta)
\end{eqnarray}

We obtain from Eq.~\ref{eq:eqm1}
\begin{widetext}
\begin{eqnarray}
\left(1-\frac{\mathbf{Y}_2\otimes\mathbf{Y}_2}{d^2}\right)
\frac{M_\perp}{\Delta^2}
\Bigl\lbrace\left(1+a_\perp\right)\mathbf{Y}_1(+)
-2\mathbf{Y}_1(0)
+\left(1-a_\perp\right)\mathbf{Y}_1(-)\Bigr\rbrace
\nonumber\\
+\left(\frac{\mathbf{Y}_2\otimes\mathbf{Y}_2}{d^2}\right)
\frac{M_\parallel}{\Delta^2}
\Bigl\lbrace\left(1+a_\parallel\right)\mathbf{Y}_1(+)
-2\mathbf{Y}_1(0)
+\left(1-a_\parallel\right)\mathbf{Y}_1(-)\Bigr\rbrace
\nonumber\\
-\frac{1}{2}\mathbf{m}^{-\frac{1}{2}}\left(\mathbf{F}_1+\mathbf{F}_2\right)
-\mathbf{G}_1
&=&0
\label{eq:discretizedeqmofy1}
\end{eqnarray}
\end{widetext}
where $a_\perp=\frac{\gamma_\perp \Delta}{2}$ and
$a_\parallel=\frac{\gamma_\parallel \Delta}{2}$. In the projectors we
have dropped the argument of $\mathbf{Y}_2(0)$.

Eq.~\ref{eq:discretizedeqmofy1} can be resolved for $\mathbf{Y}_1(+)$
by multiplication with
\begin{eqnarray}
&&\left(1-\frac{\mathbf{Y}_2\otimes\mathbf{Y}_2}{d^2}\right)
\frac{\Delta^2}{M_\perp(1+a_\perp)} \nonumber \\
&&+
\left(\frac{\mathbf{Y}_2\otimes\mathbf{Y}_2}{d^2}\right)
\frac{\Delta^2}{M_\parallel(1+a_\parallel)}
\end{eqnarray}

The result is
 \begin{widetext}
 \begin{eqnarray}
 \mathbf{Y}_1(+)&=&
 \left(1-\frac{\mathbf{Y}_2\otimes\mathbf{Y}_2}{d^2}\right)
 \Bigl\lbrace
 \frac{2}{1+a_\perp}\mathbf{Y}_1(0)
 -\frac{1-a_\perp}{1+a_\perp}\mathbf{Y}_1(-)
 +\left(\frac{1}{2}\mathbf{m}^{-\frac{1}{2}}(\mathbf{F}_1+\mathbf{F}_2)
 +\mathbf{G}_1\right)\frac{\Delta^2}{M_\perp(1+a_\perp)}\Bigr\rbrace
 \nonumber\\
 &&+\left(\frac{\mathbf{Y}_2\otimes\mathbf{Y}_2}{d^2}\right)
 \Bigl\lbrace
 \frac{2}{1+a_\parallel}\mathbf{Y}_1(0)
 -\frac{1-a_\parallel}{1+a_\parallel}\mathbf{Y}_1(-)
 +\left(\frac{1}{2}\mathbf{m}^{-\frac{1}{2}}(\mathbf{F}_1+\mathbf{F}_2)
 +\mathbf{G}_1\right)\frac{\Delta^2}{M_\parallel(1+a_\parallel)}\Bigr\rbrace
 \end{eqnarray}
 \end{widetext}
Similarly we discretize the equation of motion for the dimer distance,
Eq.~\ref{eq:eqm2}. Following Ryckaert et
al \cite{ryc77}, we first propagate without the
force of constraint to obtain 
\begin{eqnarray}
\bar{\mathbf{Y}}_2&=&
\frac{2}{1+a_\circ}\mathbf{Y}_2(0)
-\frac{1-a_\circ}{1+a_\circ}\mathbf{Y}_2(-)
\nonumber\\
&&+\left(\mathbf{m}^{-\frac{1}{2}}(\mathbf{F}_1-\mathbf{F}_2)
+\mathbf{G}_2\right)
\frac{\Delta^2}{M_\circ(1+a_\circ)}
\label{eq:discretizedeqm2}
\end{eqnarray}
with $a_\circ=\frac{\gamma_\circ\Delta}{2}$. The new vector
$\mathbf{Y}_2(+)$ is related to $\bar{Y}_2$ by the constraint force
and the Lagrange parameter $\lambda$ according to
\begin{eqnarray}
\mathbf{Y}_2(+)=\bar{\mathbf{Y}}_2
-4\lambda\mathbf{Y}_2(0)\frac{\Delta^2}{M_\circ(1+a_\circ)}
\end{eqnarray}
The Lagrange parameter is then adjusted so that the constraint
$Y_2^2(+)=d^2$, namely a given dimer length, is satisfied.

In the first iteration we estimate this forces $\mathbf{G}_1$ and
 $\mathbf{G}_2$ from the previous iterations or we set them to
 zero. Then we propagate the positions, which provides us with a
 better estimate for the nonlinear terms. This loop is then iterated
 to convergence.

\subsection{Restricting the orientation of the dimer}

It will be important to limit the degrees of freedom, in which the two
monomers may differ. This implies restricting the orientation of the
dimer to a space with lower dimensionality.  The reason is to avoid
that the dimer converges at irrelevant saddle-points, which are not of
interest. This is a common problem for complex systems, that exhibit
many minima and saddle points. This restriction is accomplished easily,
by setting the corresponding components of $\mathbf{Y}_2$ in
Eq.~\ref{eq:discretizedeqm2} to zero.

\section{\label{sec:app}
Application: Conrotatory Ring opening of 1-Chloro-2-cyclobutene}

In order to demonstrate the performance of the dimer method
described above, we have chosen a prototypical system for tests of
transition state searches, namely the ring opening of cyclobutene.
The isomerization of cyclobutene to cis-butadiene is the prototypical
example of concerted stereospecific reactions. The underlying
processes have been studied and discussed extensively by several
groups \cite{cyc1,cyc2,cyc3,cyc4} (and references therein).  The
potential energy surfaces of the con- as well as of the disrotatory
isomerization mechanisms exhibit principal structures, which render
the determination of the corresponding transition states complicated
\cite{cyc2}.

\begin{figure}[h!]
\begin{center}
\includegraphics[width=8.5cm]{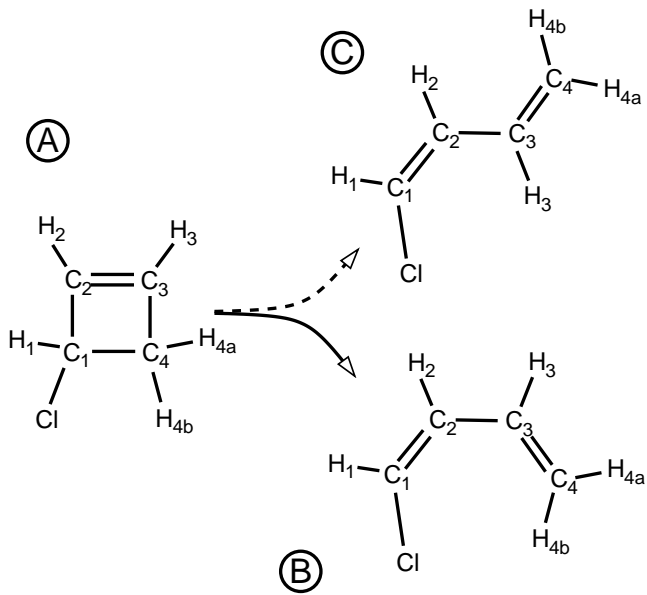}
\end{center}
\caption{Initial state, 1-chloro-2-cyclobutene (A), and final state,
  1-chloro-buta-1,3-diene (B), for the conrotatory ring opening of
  chlorocyclobutene.  Also shown is the product of the disrotatory
  ring opening, trans-1-chloro-buta-1,3-diene (C).}
\label{fig:structures1}
\end{figure}

In order to avoid special effects due to the high symmetry of
cyclobutene, that would be untypical for other molecules, we explored
1-chloro-2-cyclobutene. This molecule and the two relevant reaction
products are shown in Figure~\ref{fig:structures1}.

Computational details of our calculations are given in
appendix~\ref{sec:calcdetails}. The coordinates of the structures are
supplied with the supplementary material.

\subsection{\label{sec:dragcalc}Drag Calculations}

In this section we investigate the reaction using the drag method.
Because of its simplicity, the drag method is widely used for
transition state search. However, the drag method fails for a number
of systems. In such cases the energy exhibits a hysteresis effect,
that is, different energy profiles are obtained when dragging the
reaction coordinate in the opposite directions.  The ring opening of
cyclobutene is one example where the drag method exhibits a hysteresis
effect.

In the drag method one specifies a one-dimensional reaction
coordinate.  Then one maps the energy profile along the reaction
coordinate and determines the highest point as the transition state.
Each point along the energy profile corresponds to the local energy
minimum on a hyperplane perpendicular to the reaction coordinate. In
our case we selected the reaction coordinate by the difference between
initial and final state. A hyperplane with a specified value $c$ of
the reaction coordinate is given by
\begin{eqnarray}
\frac{\left(\mathbf{R}_{B}-\mathbf{R}_A\right)\left(\mathbf{R}-\mathbf{R}_A\right)}{\left(\mathbf{R}_{B}-\mathbf{R}_A\right)^2}=c
\end{eqnarray}
Hence the value of the reaction coordinate is zero for the initial
state (A) and one for the final state (B).

In addition to the reaction coordinate we imposed six additional
constraints to avoid translations and rotations of the molecule. We
have chosen
\begin{eqnarray}
\frac{1}{2}\left(\mathbf{R}(C_2)+\mathbf{R}(C_3)\right)=0
\\
z(C_1)=z(C_4)=\frac{1}{2}\left(x(C_1)+x(C_4)\right)=0
\end{eqnarray}
Where the coordinates correspond to the atoms given in parenthesis.
The notation follows Figure~\ref{fig:structures1}.

\begin{figure}[h!]
\begin{center}
\includegraphics[width=8.5cm]{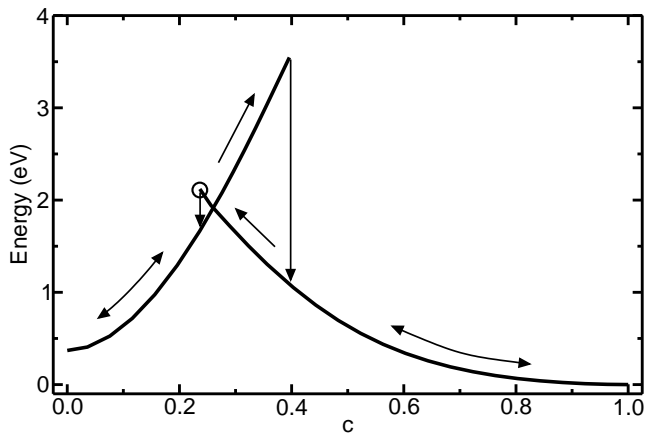}
\end{center}
\caption{Potential energy relative to the energy of
  final state (B) over reaction coordinate from a forward and backward 
calculation using the drag method. The true transition state
(determined using the dynamical dimer method) is
denoted by a circle.}
\label{fig:hyst}
\end{figure}

When the energy profile is determined by varying the reaction
coordinate in small steps, once from zero to one and then from one to
zero, we obtain the hysteresis shown in Fig.~\ref{fig:hyst}.  The
highest point of the energy profile is not the transition
state. Instead, the highest point of each branch is an upper bound for the activation energy, while the
crossing of the two branches is a lower bound.

\begin{figure}[h!]
\begin{center}
\includegraphics[width=8.5cm]{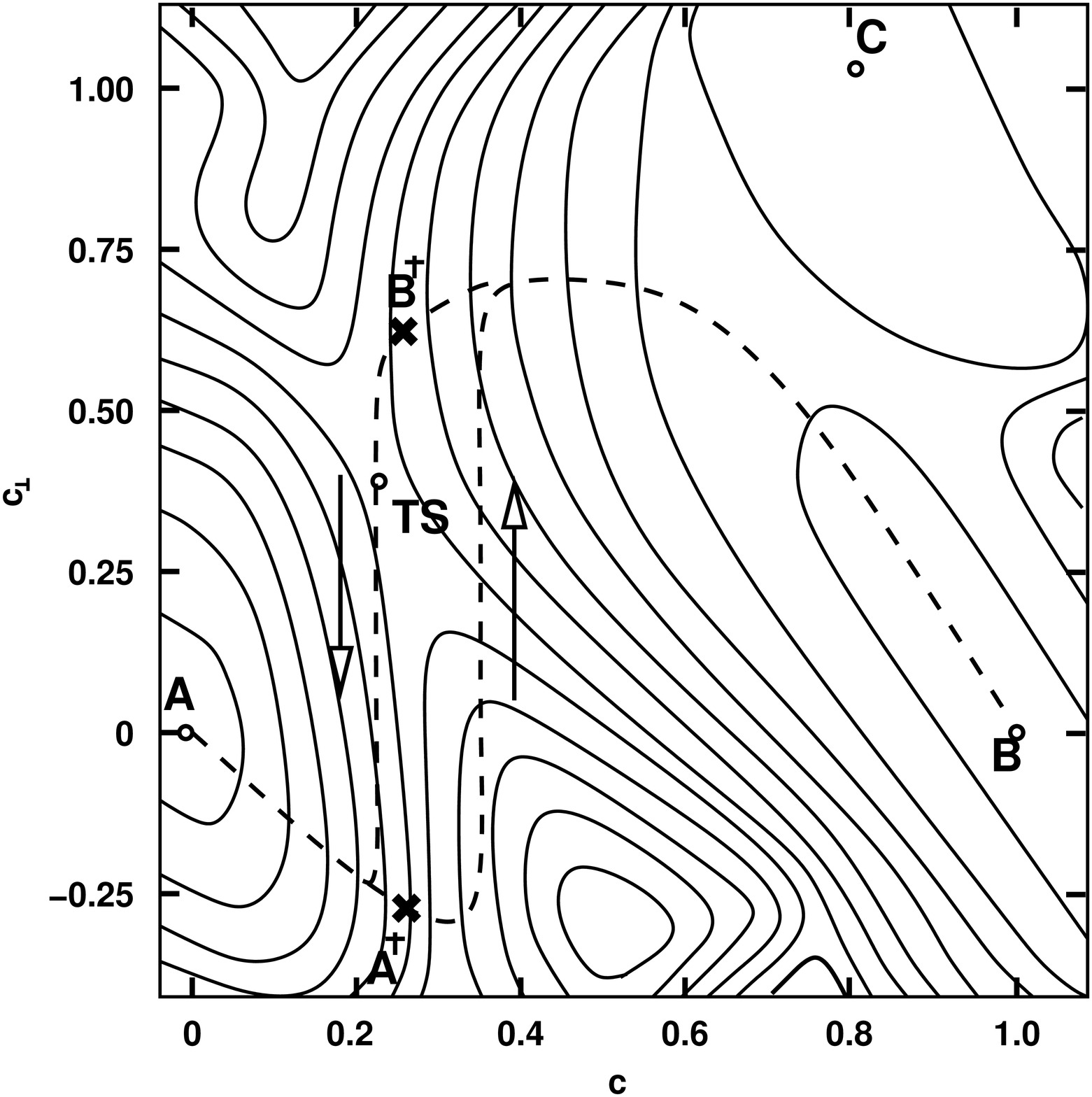}
\end{center}
\caption{Sketch of the potential energy surface of the
con-rotatory isomerization of 1-chloro-2-cyclo-butene to
1-chloro-buta-1,3-diene. The dashed lines trace the paths obtained from drag
calculations for the transitions A $\rightarrow$ B and B $\rightarrow$ A, respectively.  
The crosses denote the structures $A^\dagger$ and
$B^\dagger$. TS corresponds to the transition state.}
\label{fig:cyclopot}
\end{figure}

In order to show the underlying reason for the hysteresis, we show in
Fig.~\ref{fig:cyclopot} the total energy surface in a two-dimensional
hypersurface. The two axes are the reaction coordinate and the
coordinate defined by the two isoenergetic structures at the crossing
of the two branches of the energy profile in Fig.~\ref{fig:hyst}. For
each point in the two-dimensional plot the energy is at a local
minimum of the $(3N-2)$-dimensional hypersurface.

We see that the drag method in the forward direction leads into a side
valley that leads further onto a ridge. The point where this path becomes
unstable, and from where it leads down into the valley of the product
state, lies past the transition state. In the backward direction a
similar instability occurs, even though closer to the reaction
coordinate of the transition state. 

Note that the two configuration $A^\dagger$ and $B^\dagger$, that
mark the crossing of the two branches in the energy profile shown in
Fig.~\ref{fig:hyst}, lie far from the
transition state. For further information see Table \ref{tab:table1}.

\subsection{\label{sec:dyndimerapp}Dynamical dimer calculations}

In this section we describe technical issues for dynamical dimer
calculations and demonstrate their performance.

We use the fictitious Lagrangian formulation of ab-initio molecular
dynamics. This implies that wave function coefficients and nuclei obey
Newton's equation of motion, to which we added a friction, which
allows to quench the system.

Special attention should be given to the wave function dynamics. The
wave-function cloud tied to the atoms results in an increased
effective mass of the nuclei. Furthermore, a friction applied to the
wave function dynamics acts like an effective friction on the nuclear
motion. For a normal ground-state search, this is not problematic. In
our case however, the mass for the motion parallel to the dimer
direction is inverted. Thus the dimer accelerates opposite to the
direction of the force. Thus a friction acting on the dimer via the
electrons leads to an velocity-dependent acceleration of the dimer,
that may cause an instability of the dimer motion.

The detrimental effect of the wave function motion can be avoided by
choosing a sufficiently small mass for the wave function dynamics or by
 artificially increasing the atomic masses. In our study, we used a
mass of 50~u for all atoms. The wave functions have been kept close to
the ground state by applying a friction that dissipates 2~\% of the
kinetic energy in each time step. 

The dimension-less masses for the dimer motion have been chosen to
\begin{eqnarray}
M_{\parallel}&=&-1.00\\
M_{\perp}&=&+1.00\\
M_{\circ}&=&+0.25\ \ . 
\end{eqnarray}
The small value for $M_{\circ}$ has been chosen to speed up the
reorientation of the dimer.

In order to avoid instabilities, we found it useful to introduce an
upper limit to the kinetic energy individually for the rotational,
perpendicular and parallel motion.  These limits were enforced by
adjusting the frictions accordingly.  The limit for the kinetic energy
$E_{kin,max}$ is translated into an upper ``temperature'' $T_{max}$
according to
\begin{eqnarray}
\frac{1}{2}gk_B T_{max}=E_{kin,max}
\end{eqnarray}
where $g$ is the number of degrees of freedom in the corresponding
motion type, $k_B$ is Boltzmann's constant.

Within the limits of the enforced maximum kinetic energies, we adapted
the friction dynamically in each step to come close to the limit of
critical damping. The latter results in the best possible convergence
behavior.

Furthermore, the friction is increased to a higher value specified by
the user, if the dimer moves away from the stable point. The decision
to increase the friction has been made on the basis of the actual
forces and velocities. In our simulation we have chosen this friction
so that the energy dissipated per time step corresponds to 20~\% of
the kinetic energy of the corresponding type of motion.

When starting the dimer simulation, it is important to optimize the
dimer orientation, i.e. $\mathbf{Y}_1$, first, before the mean dimer
configuration $\mathbf{Y}_2$ changes appreciably. This is because the
mean dimer configuration only approaches the transition state, if the
dimer orientation is sufficiently aligned along the unstable
vibrational mode of the transition state.

The dimer orientation can be optimized in different ways. (1) Starting
from initial dimer coordinates, the mean dimer configuration, $\mathbf{Y}_1$, is
constrained to the initial value, while the orientation, $\mathbf{Y}_2$, is fully
optimized. Only after this optimization also the mean configuration is
allowed to move. (2) Starting from one monomer configuration, the
second monomer is constructed from the first by letting it follow the
forces until the desired dimer length is obtained. We will refer to
this technique as the ``growing-dimer technique'' (3) Starting from a
dimer with the monomers being identical to the two metastable minima,
the dimer length is successively shortened, while the dimer position
is optimized. Results for all three optimization strategies can be
found in Table~\ref{tab:table1}.

We found that the third strategy suffers from the fact that additional
effort is required to contract the dimer length to the desired value.
It also suffers from an instability caused by strong anharmonic
effects. 

The growing-dimer technique has the advantage that the second monomer
is constructed dynamically from the first, without the need of an
independent optimization of the wave functions.  In the growing dimer
technique the dimer orientation is optimized automatically as soon as
the targeted dimer length is reached. In the first strategy the
orientation is obtained during the first iterations with a constrained
mean dimer configuration.

We will discuss here the first strategy. The initial structure of the
dimer is given by a mean dimer configuration midway between the two
metastable states (A) and (B). The dimer length is chosen such that the
distance of the two monomers in coordinate space is 0.33~\AA.

\begin{figure}[h!]
\begin{center}
\includegraphics[width=8.5cm]{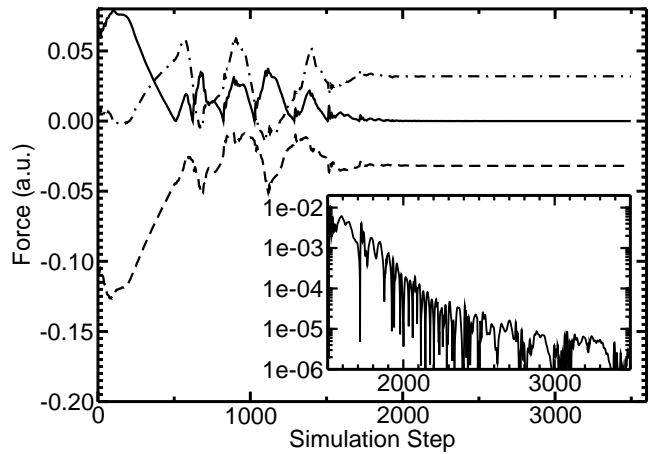}
\end{center}
\caption{Forces over time-step for a calculation following optimzation
  strategy (1) described in
  the text. The dashed and dash-dotted line show the
forces parallel to the dimer axis for image one and two,
respectively. The full line corresponds to the resulting force acting on the
center of gravity of the dimer. The logarithmic scale of the inset
provides detailed information for the latter.}
\label{fig:forces1}
\end{figure}

\begin{figure}[h!]
\begin{center}
\includegraphics[width=8.5cm]{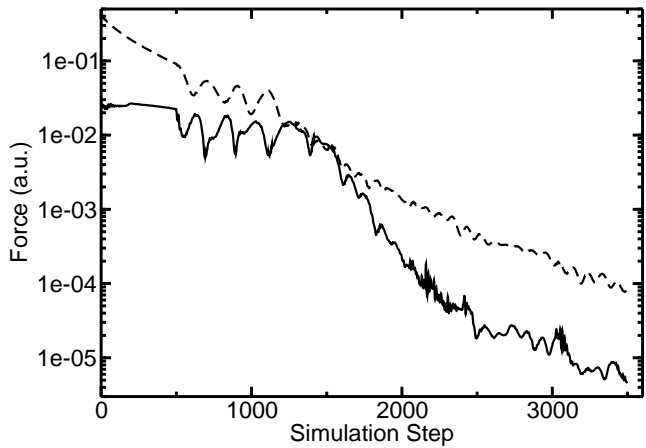}
\end{center}
\caption{Rotational (full line) and perpendicular (dashed line) part of the forces acting on the
dimer over time-steps for a calculation following optimzation strategy (1) described in the text.}
\label{fig:forces2}
\end{figure}

During the first 700 time steps, the dimer orientation has been
optimized. Here the mean dimer configuration has not been constrained,
but the maximum temperature for the perpendicular motion has been kept
at a small value of 10~K. For the parallel and rotational motion, the maximum
temperature has been set to 500~K. After the first 700 time steps the
maximum temperature for the perpendicular motion has been increased to
500~K. In Figs.~\ref{fig:forces1},\ref{fig:forces2} we show the
convergence of the forces for the parallel, rotational and
perpendicular motion. Good convergence is reached after 2000 time
steps. The estimates for the transition state energy and the deviation
of the mean dimer configuration from the transition state is given in
Table~\ref{tab:table1}.


\begin{table}
\caption{\label{tab:table1}Predicted transition-state energy (relative
  to final state (B)) and
  deviation from the exact transition state for the con-rotatory
  isomerization of 1-chloro-2-cyclo-butene to
  1-chloro-buta-1,3-diene. The results include three different
  dynamical dimer optimization strategies ((1), (2) and (3)) described
  in the text and the forward as well as the backward drag
  calculation shown in Fig.~\ref{fig:hyst}.
}
\begin{ruledtabular}
\begin{tabular}{ccccc}
 &&Predicted Energy (eV)&$\left|\mathbf{R}^{TS}-\mathbf{R}\right|$ (\AA)&\\
\hline
\multicolumn{2}{c}{Transition State\footnotemark[1]} &2.110 &- & \\
\hline
(1)&  &  &  \\
&\ 1000 steps&    2.110  &1.00  &  \\
&\ 1500 steps&    2.111  &0.85  &  \\
&\ 2000 steps&    2.111  &0.71  &  \\
(2)&  &  &  \\
&\ 1000 steps&    2.258  &0.62  &  \\
&\ 1500 steps&    2.116  &0.37  &  \\
&\ 2000 steps&    2.110  &0.36  &  \\
(3)&  &  &  \\
&\ 1000 steps&    2.133  &1.58  &  \\
&\ 1500 steps&    2.112  &0.85  &  \\
&\ 2000 steps&    2.110  &0.85  &  \\
\multicolumn{2}{c}{TS $A\rightarrow B$} &    3.461 &2.57 & \\
\multicolumn{2}{c}{TS $B\rightarrow A$} &    2.120 &0.56 & \\
\end{tabular}
\end{ruledtabular}
\footnotetext[1]{Determined by the use of a well converged dynamical
  dimer calculation with final dimer distance of
  0.125~\AA.}
\end{table}

\section{\label{sec:concl}Summary}

A formulation of the dimer method for searching transition states
is  presented  which  can  be  used in  ab-initio  molecular  dynamics
simulations  using  a  fictitious  Lagrangian.  The  dimer  method  is
successful  in cases  where  the  widely used  drag  method fails.  We
investigate  the dynamics  close to  the  stable points  of the  dimer
analytically.  In  addition  we  demonstrate the  performance  of  our
implementation using  a practical example typically used  as test case
for transition  state search  algorithms, namely the con-rotatory ring
opening of (chloro-) cyclo-butene.

\appendix
\section{\label{sec:calcdetails}Computational details}

We performed density-functional calculations \cite{hoh64,koh65} based
on the projector augmented wave (PAW) method\cite{blo94,blo03}. The
gradient-corrected PBE\cite{per96} functional was used for exchange
and correlation.  The PAW method is a frozen-core all-electron method.
Like other plane-wave based methods, the PAW method leads to the
occurrence of artificial periodic images of the structures. This
effect was avoided by explicit subtraction of the electrostatic
interaction between them.\cite{blo86} Wave function overlap was
avoided by choosing the unit cell large enough to keep a distance of
more than 6\,\AA{} between atoms belonging to different periodic
images. We used a plane wave cutoff of 30\,Ry for the auxiliary wave
functions of the PAW method. The following sets of projector functions
were employed, Cl 2s2p1d, C 2s2p1d, H 2s1p, which provides the number
of projector functions per angular momentum magnetic quantum number
$m$ in each main angular momentum channel $\ell$.

Atomic structures were optimized by damped Car-Parrinello\cite{car85}
molecular dynamics. We used a time-step of 10\,a.u. (2.5\,fs) for all
except the dynamical dimer calculations. See section
\ref{sec:dyndimerapp} for further details. The convergence was tested by monitoring if the total
energy change remains below $10^{-5}$\,Hartree during a simulation of
500 time steps.  During the simulation for the convergence test, no friction was applied to
the atomic motion and the friction on the wave function dynamics was
chosen sufficiently low to avoid a noticeable effect on the atomic
motion.

\newpage 
\bibliography{apssamp}

\begin{thebibliography}{33}
\expandafter\ifx\csname natexlab\endcsname\relax\def\natexlab#1{#1}\fi
\expandafter\ifx\csname bibnamefont\endcsname\relax
  \def\bibnamefont#1{#1}\fi
\expandafter\ifx\csname bibfnamefont\endcsname\relax
  \def\bibfnamefont#1{#1}\fi
\expandafter\ifx\csname citenamefont\endcsname\relax
  \def\citenamefont#1{#1}\fi
\expandafter\ifx\csname url\endcsname\relax
  \def\url#1{\texttt{#1}}\fi
\expandafter\ifx\csname urlprefix\endcsname\relax\def\urlprefix{URL }\fi
\providecommand{\bibinfo}[2]{#2}
\providecommand{\eprint}[2][]{\url{#2}}

\bibitem[{\citenamefont{Eyring}(1934)}]{ts1}
\bibinfo{author}{\bibfnamefont{H.}~\bibnamefont{Eyring}}, \bibinfo{journal}{J.\
  Chem.\ Phys.} \textbf{\bibinfo{volume}{3}}, \bibinfo{pages}{107}
  (\bibinfo{year}{1934}).

\bibitem[{\citenamefont{Vineyard}(1957)}]{ts2}
\bibinfo{author}{\bibfnamefont{G.}~\bibnamefont{Vineyard}},
  \bibinfo{journal}{J.\ Phys.\ Chem.\ Solids} \textbf{\bibinfo{volume}{3}},
  \bibinfo{pages}{121} (\bibinfo{year}{1957}).

\bibitem[{\citenamefont{Keck}(1962)}]{keck62}
\bibinfo{author}{\bibfnamefont{J.}~\bibnamefont{Keck}},
  \bibinfo{journal}{Discuss. Faraday Soc.} \textbf{\bibinfo{volume}{33}},
  \bibinfo{pages}{1962} (\bibinfo{year}{1962}).

\bibitem[{\citenamefont{Poppinger}(1975)}]{fts1}
\bibinfo{author}{\bibfnamefont{D.}~\bibnamefont{Poppinger}},
  \bibinfo{journal}{Chem.\ Phys.\ Lett.} \textbf{\bibinfo{volume}{35}},
  \bibinfo{pages}{550} (\bibinfo{year}{1975}).

\bibitem[{\citenamefont{Cerjan and Miller}(1981)}]{fts2}
\bibinfo{author}{\bibfnamefont{C.~J.} \bibnamefont{Cerjan}} \bibnamefont{and}
  \bibinfo{author}{\bibfnamefont{W.~H.} \bibnamefont{Miller}},
  \bibinfo{journal}{J.\ Chem.\ Phys.} \textbf{\bibinfo{volume}{75}},
  \bibinfo{pages}{2800} (\bibinfo{year}{1981}).

\bibitem[{\citenamefont{Baker}(1986)}]{fts3}
\bibinfo{author}{\bibfnamefont{J.}~\bibnamefont{Baker}}, \bibinfo{journal}{J.\
  Comp.\ Chem.} \textbf{\bibinfo{volume}{7}}, \bibinfo{pages}{385}
  (\bibinfo{year}{1986}).

\bibitem[{\citenamefont{Simons et~al.}(1983)\citenamefont{Simons, Jorgensen,
  Taylor, and Ozment}}]{fts4}
\bibinfo{author}{\bibfnamefont{J.}~\bibnamefont{Simons}},
  \bibinfo{author}{\bibfnamefont{P.}~\bibnamefont{Jorgensen}},
  \bibinfo{author}{\bibfnamefont{H.}~\bibnamefont{Taylor}}, \bibnamefont{and}
  \bibinfo{author}{\bibfnamefont{J.}~\bibnamefont{Ozment}},
  \bibinfo{journal}{J.\ Phys.\ Chem.} \textbf{\bibinfo{volume}{87}},
  \bibinfo{pages}{2745} (\bibinfo{year}{1983}).

\bibitem[{\citenamefont{Banerjee et~al.}(1985)\citenamefont{Banerjee, Adams,
  Simons, and Shepard}}]{fts5}
\bibinfo{author}{\bibfnamefont{A.}~\bibnamefont{Banerjee}},
  \bibinfo{author}{\bibfnamefont{N.}~\bibnamefont{Adams}},
  \bibinfo{author}{\bibfnamefont{J.}~\bibnamefont{Simons}}, \bibnamefont{and}
  \bibinfo{author}{\bibfnamefont{R.}~\bibnamefont{Shepard}},
  \bibinfo{journal}{J.\ Phys.\ Chem.} \textbf{\bibinfo{volume}{89}},
  \bibinfo{pages}{52} (\bibinfo{year}{1985}).

\bibitem[{\citenamefont{Munro and Wales}(1999)}]{fts6}
\bibinfo{author}{\bibfnamefont{L.~J.} \bibnamefont{Munro}} \bibnamefont{and}
  \bibinfo{author}{\bibfnamefont{D.~J.} \bibnamefont{Wales}},
  \bibinfo{journal}{Phys. Rev. B} \textbf{\bibinfo{volume}{59}},
  \bibinfo{pages}{3969} (\bibinfo{year}{1999}).

\bibitem[{\citenamefont{Jonsson et~al.}(1998)\citenamefont{Jonsson, Mills, and
  Jacobsen}}]{fts7}
\bibinfo{author}{\bibfnamefont{H.}~\bibnamefont{Jonsson}},
  \bibinfo{author}{\bibfnamefont{G.}~\bibnamefont{Mills}}, \bibnamefont{and}
  \bibinfo{author}{\bibfnamefont{K.~W.} \bibnamefont{Jacobsen}},
  \bibinfo{journal}{\textit{Classical and Quantum Dynamics in Condensed Phase
  Simulations}, Ed. B.J. Berne, G.Ciccotti and D.F.Coker, World Scientific} p.
  \bibinfo{pages}{385} (\bibinfo{year}{1998}).

\bibitem[{\citenamefont{Mills and Jonsson}(1994)}]{fts8}
\bibinfo{author}{\bibfnamefont{G.}~\bibnamefont{Mills}} \bibnamefont{and}
  \bibinfo{author}{\bibfnamefont{H.}~\bibnamefont{Jonsson}},
  \bibinfo{journal}{Phys.\ Rev.\ Lett.} \textbf{\bibinfo{volume}{72}},
  \bibinfo{pages}{1124} (\bibinfo{year}{1994}).

\bibitem[{\citenamefont{Mills et~al.}(1995)\citenamefont{Mills, H.Jonsson, and
  Schenter}}]{fts9}
\bibinfo{author}{\bibfnamefont{G.}~\bibnamefont{Mills}},
  \bibinfo{author}{\bibnamefont{H.Jonsson}}, \bibnamefont{and}
  \bibinfo{author}{\bibfnamefont{G.}~\bibnamefont{Schenter}},
  \bibinfo{journal}{Surf.\ Sci.} \textbf{\bibinfo{volume}{324}},
  \bibinfo{pages}{305} (\bibinfo{year}{1995}).

\bibitem[{\citenamefont{Malek and Mousseau}(2000)}]{fts10}
\bibinfo{author}{\bibfnamefont{R.}~\bibnamefont{Malek}} \bibnamefont{and}
  \bibinfo{author}{\bibfnamefont{N.}~\bibnamefont{Mousseau}},
  \bibinfo{journal}{Phys.\ Rev.\ E} \textbf{\bibinfo{volume}{62}},
  \bibinfo{pages}{7723} (\bibinfo{year}{2000}).

\bibitem[{\citenamefont{E et~al.}(2002)\citenamefont{E, Ren, and
  Vanden-Eijnden}}]{fts11}
\bibinfo{author}{\bibfnamefont{W.}~\bibnamefont{E}},
  \bibinfo{author}{\bibfnamefont{W.}~\bibnamefont{Ren}}, \bibnamefont{and}
  \bibinfo{author}{\bibfnamefont{E.}~\bibnamefont{Vanden-Eijnden}},
  \bibinfo{journal}{Phys.\ Rev.\ B} \textbf{\bibinfo{volume}{66}},
  \bibinfo{pages}{52301} (\bibinfo{year}{2002}).

\bibitem[{\citenamefont{Peters et~al.}(2004)\citenamefont{Peters, Heyden, Bell,
  and Chakraborty}}]{fts12}
\bibinfo{author}{\bibfnamefont{B.}~\bibnamefont{Peters}},
  \bibinfo{author}{\bibfnamefont{A.}~\bibnamefont{Heyden}},
  \bibinfo{author}{\bibfnamefont{A.~T.} \bibnamefont{Bell}}, \bibnamefont{and}
  \bibinfo{author}{\bibfnamefont{A.}~\bibnamefont{Chakraborty}},
  \bibinfo{journal}{J.\ Chem.\ Phys.} \textbf{\bibinfo{volume}{120}},
  \bibinfo{pages}{7877} (\bibinfo{year}{2004}).

\bibitem[{\citenamefont{Henkelman et~al.}(2000)\citenamefont{Henkelman,
  Johannesson, and Jonsson}}]{ftsover}
\bibinfo{author}{\bibfnamefont{G.}~\bibnamefont{Henkelman}},
  \bibinfo{author}{\bibfnamefont{G.}~\bibnamefont{Johannesson}},
  \bibnamefont{and} \bibinfo{author}{\bibfnamefont{H.}~\bibnamefont{Jonsson}},
  \bibinfo{journal}{\textit{Progress on Theoretical Chemistry and Physics}, Ed.
  S. D. Schwartz, Kluwer Academic Publishers} p. \bibinfo{pages}{269}
  (\bibinfo{year}{2000}).

\bibitem[{\citenamefont{Schlegel}(2003)}]{ftsover2}
\bibinfo{author}{\bibfnamefont{H.~B.} \bibnamefont{Schlegel}},
  \bibinfo{journal}{J.\ Comput.\ Chem.} \textbf{\bibinfo{volume}{24}},
  \bibinfo{pages}{1514} (\bibinfo{year}{2003}).

\bibitem[{\citenamefont{Henkelman and Jonsson}(1999)}]{dimer1}
\bibinfo{author}{\bibfnamefont{G.}~\bibnamefont{Henkelman}} \bibnamefont{and}
  \bibinfo{author}{\bibfnamefont{H.}~\bibnamefont{Jonsson}},
  \bibinfo{journal}{J.\ Chem.\ Phys.} \textbf{\bibinfo{volume}{111}},
  \bibinfo{pages}{7010} (\bibinfo{year}{1999}).

\bibitem[{\citenamefont{Olsen et~al.}(2004)\citenamefont{Olsen, Kroes,
  Henkelman, Arnoldsson, and H.Jonsson}}]{dimer2}
\bibinfo{author}{\bibfnamefont{R.~A.} \bibnamefont{Olsen}},
  \bibinfo{author}{\bibfnamefont{G.~H.} \bibnamefont{Kroes}},
  \bibinfo{author}{\bibfnamefont{G.}~\bibnamefont{Henkelman}},
  \bibinfo{author}{\bibfnamefont{A.}~\bibnamefont{Arnoldsson}},
  \bibnamefont{and} \bibinfo{author}{\bibnamefont{H.Jonsson}},
  \bibinfo{journal}{J.\ Chem.\ Phys.} \textbf{\bibinfo{volume}{121}},
  \bibinfo{pages}{9776} (\bibinfo{year}{2004}).

\bibitem[{\citenamefont{Car and Parrinello}(1985{\natexlab{a}})}]{cp}
\bibinfo{author}{\bibfnamefont{R.}~\bibnamefont{Car}} \bibnamefont{and}
  \bibinfo{author}{\bibfnamefont{M.}~\bibnamefont{Parrinello}},
  \bibinfo{journal}{Phys.\ Rev.\ Lett.} \textbf{\bibinfo{volume}{55}},
  \bibinfo{pages}{2471} (\bibinfo{year}{1985}{\natexlab{a}}).

\bibitem[{\citenamefont{Kanai et~al.}(2004)\citenamefont{Kanai, Tilocca,
  Selloni, and Car}}]{stringcar}
\bibinfo{author}{\bibfnamefont{Y.}~\bibnamefont{Kanai}},
  \bibinfo{author}{\bibfnamefont{A.}~\bibnamefont{Tilocca}},
  \bibinfo{author}{\bibfnamefont{A.}~\bibnamefont{Selloni}}, \bibnamefont{and}
  \bibinfo{author}{\bibfnamefont{R.}~\bibnamefont{Car}}, \bibinfo{journal}{J.\
  Chem.\ Phys.} \textbf{\bibinfo{volume}{121}}, \bibinfo{pages}{3359}
  (\bibinfo{year}{2004}).

\bibitem[{\citenamefont{Ryckaert et~al.}(1977)\citenamefont{Ryckaert, Ciccotti,
  and Berendsen}}]{ryc77}
\bibinfo{author}{\bibfnamefont{J.-P.} \bibnamefont{Ryckaert}},
  \bibinfo{author}{\bibfnamefont{G.}~\bibnamefont{Ciccotti}}, \bibnamefont{and}
  \bibinfo{author}{\bibfnamefont{H.~J.~C.} \bibnamefont{Berendsen}},
  \bibinfo{journal}{J. Comp. Phys.} \textbf{\bibinfo{volume}{23}},
  \bibinfo{pages}{327} (\bibinfo{year}{1977}).

\bibitem[{\citenamefont{Woodward and Hoffmann}(1965)}]{cyc1}
\bibinfo{author}{\bibfnamefont{R.~B.} \bibnamefont{Woodward}} \bibnamefont{and}
  \bibinfo{author}{\bibfnamefont{R.}~\bibnamefont{Hoffmann}},
  \bibinfo{journal}{J. \ Am. \ Chem. \ Soc.} \textbf{\bibinfo{volume}{87}},
  \bibinfo{pages}{395} (\bibinfo{year}{1965}).

\bibitem[{\citenamefont{Dewar and Kirschner}(1971)}]{cyc2}
\bibinfo{author}{\bibfnamefont{M.~J.~S.} \bibnamefont{Dewar}} \bibnamefont{and}
  \bibinfo{author}{\bibfnamefont{S.}~\bibnamefont{Kirschner}},
  \bibinfo{journal}{J. \ Am. \ Chem. \ Soc.} \textbf{\bibinfo{volume}{93}},
  \bibinfo{pages}{4292} (\bibinfo{year}{1971}).

\bibitem[{\citenamefont{Breulet and Schaefer}(1981)}]{cyc3}
\bibinfo{author}{\bibfnamefont{J.}~\bibnamefont{Breulet}} \bibnamefont{and}
  \bibinfo{author}{\bibfnamefont{H.~F.} \bibnamefont{Schaefer}},
  \bibinfo{journal}{J. \ Am. \ Chem. \ Soc.}  (\bibinfo{year}{1981}).

\bibitem[{\citenamefont{Olivia et~al.}(1997)\citenamefont{Olivia, Gerratt,
  Karadakov, and Cooper}}]{cyc4}
\bibinfo{author}{\bibfnamefont{J.~M.} \bibnamefont{Olivia}},
  \bibinfo{author}{\bibfnamefont{J.}~\bibnamefont{Gerratt}},
  \bibinfo{author}{\bibfnamefont{P.~B.} \bibnamefont{Karadakov}},
  \bibnamefont{and} \bibinfo{author}{\bibfnamefont{D.~L.}
  \bibnamefont{Cooper}}, \bibinfo{journal}{J. \ Chem. \ Phys.}
  \textbf{\bibinfo{volume}{107}}, \bibinfo{pages}{8917} (\bibinfo{year}{1997}).

\bibitem[{\citenamefont{Hohenberg and Kohn}(1964)}]{hoh64}
\bibinfo{author}{\bibfnamefont{P.}~\bibnamefont{Hohenberg}} \bibnamefont{and}
  \bibinfo{author}{\bibfnamefont{W.}~\bibnamefont{Kohn}},
  \bibinfo{journal}{Phys. Rev.} \textbf{\bibinfo{volume}{136}},
  \bibinfo{pages}{B864} (\bibinfo{year}{1964}).

\bibitem[{\citenamefont{Kohn and Sham}(1965)}]{koh65}
\bibinfo{author}{\bibfnamefont{W.}~\bibnamefont{Kohn}} \bibnamefont{and}
  \bibinfo{author}{\bibfnamefont{L.}~\bibnamefont{Sham}},
  \bibinfo{journal}{Phys. Rev.} \textbf{\bibinfo{volume}{140}},
  \bibinfo{pages}{A1133} (\bibinfo{year}{1965}).

\bibitem[{\citenamefont{Bl\"ochl}(1994)}]{blo94}
\bibinfo{author}{\bibfnamefont{P.~E.} \bibnamefont{Bl\"ochl}},
  \bibinfo{journal}{Phys. Rev. B} \textbf{\bibinfo{volume}{50}},
  \bibinfo{pages}{17953} (\bibinfo{year}{1994}).

\bibitem[{\citenamefont{Bl{\"o}chl et~al.}(2003)\citenamefont{Bl{\"o}chl,
  F{\"o}rst, and Schimpl}}]{blo03}
\bibinfo{author}{\bibfnamefont{P.~E.} \bibnamefont{Bl{\"o}chl}},
  \bibinfo{author}{\bibfnamefont{C.}~\bibnamefont{F{\"o}rst}},
  \bibnamefont{and} \bibinfo{author}{\bibfnamefont{J.}~\bibnamefont{Schimpl}},
  \bibinfo{journal}{Bull. Mater. Sci.} \textbf{\bibinfo{volume}{26}},
  \bibinfo{pages}{33} (\bibinfo{year}{2003}).

\bibitem[{\citenamefont{Perdew et~al.}(1996)\citenamefont{Perdew, Burke, and
  Ernzerhof}}]{per96}
\bibinfo{author}{\bibfnamefont{J.}~\bibnamefont{Perdew}},
  \bibinfo{author}{\bibfnamefont{K.}~\bibnamefont{Burke}}, \bibnamefont{and}
  \bibinfo{author}{\bibfnamefont{M.}~\bibnamefont{Ernzerhof}},
  \bibinfo{journal}{Phys. Rev. Lett.} \textbf{\bibinfo{volume}{77}},
  \bibinfo{pages}{3865} (\bibinfo{year}{1996}).

\bibitem[{\citenamefont{Bl\"ochl}(1986)}]{blo86}
\bibinfo{author}{\bibfnamefont{P.~E.} \bibnamefont{Bl\"ochl}},
  \bibinfo{journal}{J. Chem. Phys.} \textbf{\bibinfo{volume}{103}},
  \bibinfo{pages}{7422} (\bibinfo{year}{1986}).

\bibitem[{\citenamefont{Car and Parrinello}(1985{\natexlab{b}})}]{car85}
\bibinfo{author}{\bibfnamefont{R.}~\bibnamefont{Car}} \bibnamefont{and}
  \bibinfo{author}{\bibfnamefont{M.}~\bibnamefont{Parrinello}},
  \bibinfo{journal}{Phys. Rev. Lett.} \textbf{\bibinfo{volume}{55}},
  \bibinfo{pages}{2471} (\bibinfo{year}{1985}{\natexlab{b}}).

\end{thebibliography}

\end{document}